\documentclass[11pt]{article}
\usepackage[dvips]{epsfig}
\usepackage{bm,amsfonts,nicefrac}

\newtheorem{theorem}{Theorem}

\newtheorem{lemma}{Lemma}

\def\cqfd{\hfill\hbox{$\hbox{\vrule width 0.8pt
\vbox to6pt{\hrule depth 0.8pt width 5.2pt
\vfill\hrule depth 0.8pt}\vrule width 0.8pt}$}}
\def\Beta{\mathrm{B}}

\newcommand{\defeq}{\stackrel{\mathrm{def}}{=}}

\newcommand{\es}{\mbox{\textbf{E}}}
\newcommand{\pr}{\mbox{\textbf{Pr}}}

\title{Staring at Economic Aggregators through Information Lenses}

\author{Richard Nock\\
Centre d'Etude et de Recherche en Economie, Gestion, Mod\'elisation\\
et Informatique Appliqu\'ee (\textsc{Ceregmia} --- UAG),\\
PO Box 7209, Schoelcher 97275, France.\\
\texttt{rnock@martinique.univ-ag.fr}
\and
Nicolas Sanz\\
\textsc{Ceregmia} --- UAG,\\
PO Box 792, Cayenne 97400, France.\\
\texttt{Fred.Celimene@martinique.univ-ag.fr}
\and
Fred C{\'e}lim{\`e}ne\\
\textsc{Ceregmia} --- UAG,\\
PO Box 7209, Schoelcher 97275, France.\\
\texttt{Fred.Celimene@martinique.univ-ag.fr}
\and
Frank Nielsen\\
LIX --- Ecole Polytechnique, Palaiseau 91128, France\\
$\&$ Sony Computer Science Laboratories Inc., 3-14-13\\
Higashi Gotanda, Shinagawa-Ku, 141-0022 Tokyo, Japan.\\
\texttt{Nielsen@acm.org}}

\begin{document}
\maketitle

\begin{abstract}
It is hard to exaggerate the role of economic aggregators --- functions that
summarize numerous and / or heterogeneous data --- in economic models 
since the early XX$^{th}$ century. In many cases, as witnessed
by the pioneering works of Cobb and Douglas, these functions were information
quantities tailored to economic theories, i.e. they were built to fit economic
phenomena. In this paper, we look at these functions from the complementary side: information. 
We use a recent toolbox built on top
of a vast class of distortions coined by Bregman, whose application field rivals metrics'
in various subfields of mathematics. This toolbox makes it possible to find 
the quality of an aggregator (for consumptions, prices, labor, capital, wages, etc.), from the standpoint 
of the information it carries. We prove
a rather striking result. From the informational standpoint, well-known economic aggregators
 do belong to the \textit{optimal} set. 
As common economic assumptions enter the analysis, this large set shrinks, and it
essentially ends up \textit{exactly fitting} either CES, or Cobb-Douglas, or both.
To summarize, in the relevant economic contexts, one could not have crafted better some aggregator
from the information standpoint. We also discuss global economic behaviors
of optimal information aggregators in general, and present a brief panorama of 
the links between economic and information aggregators.
\end{abstract}

\section{Introduction}


Since the end of the XIX$^{th}$ century and the birth of the ``neo-classical'' school, 
mathematics have played a growing role in economics. With the works of L\'eon Walras, 
the question of aggregation of the behavior of many individuals has risen and become 
central in the economic theory. In order to represent as well as possible the evolution 
of these aggregate variables, some mathematical functions have been proposed and become 
very famous in the economic literature.

One of the most famous neo-classical function is the Cobb-Douglas \cite{cdAT,tDI}. This 
function is of particular interest, since it allows for perfect substitutability between the 
different inputs it depends on. Another well-known ``linear'' function was later formulated 
by Leontief \cite{lTS}, in which inputs are conversely complementary. The choice of such a function 
to describe the production process has very strong implications at the macroeconomic level, as 
illustrated by many results found by Keynesians economics in the literature on growth theory.

But beyond these different aggregate functions, one of the most recently built and well-known one is 
the constant elasticity of substitution (CES) function elaborated by Arrow et al. \cite{acmsCL}. 
Indeed, in the Cobb-Douglas production function, the elasticity of substitution of capital for 
labor is fixed to unity. This implies that a one percent increase in the capital stock implies an 
equal one percent fall in labor inputs in order to maintain a constant production level, given the 
structure of relative prices. On the contrary, the CES function allows this elasticity to lie between 
zero and infinity, but to stay fixed at that number along and across the isoquants, whatever the 
quantities of inputs that are used in the production process. The main advantage exhibited by the 
CES function is that it encompasses the Cobb-Douglas, the Leontief and the Linear production 
functions, which are in fact limit and thus particular cases of it. Nevertheless, one of the 
reasons economists have kept on using simpler functions such as the Cobb-Douglas one is the 
heavy calculus to which the CES function often leads, especially at the point where models have to be closed.

In a seminal work, Douglas in \cite{dTT} highlights the importance of the progresses in the field of statistical 
information in the genesis of his essay. Pioneering works of Cobb and Douglas \cite{cdAT}, and Arrow \textit{et al.}
\cite{acmsCL}, underline the inductive nature of the inception of their respective functions, as the
purpose was to fit as best as possible information quantities (aggregators) to observed economic phenomena. In this paper, we take a deductive route paved with a rigorous information material, to derive these fundamental quantities based on two assumptions:
\begin{itemize}
\item an aggregator should always be as informative as possible with respect to the data it summarizes (prices, 
consumptions, wages, capital, labor, etc.);
\item an aggregator might be require to satisfy standard economic assumptions, relying
on aggregator dualities (prices / consumptions, wages / labor, etc.), elasticities, marginal rates of substitutions,
returns to scale, etc.
\end{itemize}
The starting point of our work is a class of distortions coined in the sixties by Bregman \cite{bTR}, in
the context of convex programming. Though they were born four decades ago, it was only much later that 
these distortions literally spread out to other fields, including statistics, signal processing and 
classification \cite{gwPR}, fields where they had to become undeniably central. It was even later 
that was discovered their broad applicability, with an 
axiomatization that makes it possible to relate
them to metrics and their spawns \cite{bgwOT}. Very roughly, \textit{Bregman divergences} are non-negative
functions that meet the same identity of indiscernibles condition as metrics, and rely on a third assumption about the
existence of a particular aggregator which minimizes the total distortion to a set. This last condition,
which can be rephrased as a maximum likelihood condition, makes this aggregator the \textit{most informative} 
quantity about the data, and we call it a \textit{Low Distortion Aggregator} (LDA). 

In this paper, our contribution is threefold. First, we make a clear partition of economic aggregators
with respect to information, as we show that some \textit{are} LDAs (CES, Cobb-Douglas), some
are limit cases of LDAs (Leontief), and some are neither (Mitscherlich-Spillman-von Th{\"u}nen).
Without more assumptions, the set of all LDAs is huge, yet we show that global trends of economic
relevance can be easily shown for all, such as on marginal rates of substitution, and the set can be quite 
easily drilled down for aggregators with general behaviors, such as concavity or convexity. This, in fact, is
our last contribution. Our main contribution is to show that, when we plug in various standard economic
assumptions (see above), the set of all LDAs reduces to a particular subset which \textit{precisely} matches
CES, Cobb-Douglas, or both sets. This novel advocacy for the use of these popular aggregators
 brings a very strong information-theoretic rationale to their ``economic'' existence.

The remaining of the paper is structured as follows. Section 2 presents LDAs and their main properties. In Section 3, we relate common economic aggregators to LDAs. Section 4 discusses additional properties of LDAs. A last section concludes the paper, with avenues for future research. In order not to laden the paper's body, all proofs have been postponed to an appendix.

\section{Low-distortion aggregators}\label{secb}

For any strictly convex function $\varphi : {\mathbb{X}} \rightarrow {\mathbb{R}}$ differentiable on ${\mathrm{int}}({\mathbb{X}})$, with ${\mathbb{X}} \subseteq {\mathbb{R}}^d$ convex, the \textit{Bregman Divergence} $D_\varphi$ with generator $\varphi$ is \cite{bTR,bgwOT}:
\begin{eqnarray}
D_\varphi(\bm{x} || \bm{y})  & \defeq & \varphi(\bm{x}) - \varphi(\bm{y}) - \langle \bm{x} - \bm{y}, \bm{\nabla}_\varphi (\bm{y})\rangle \:\:, \label{bdir}
\end{eqnarray}
where $\langle\cdot,\cdot\rangle$ denotes the inner product, and $\bm{\nabla}_\varphi \defeq \left[ \partial \varphi/\partial x_i \right]^\top$ is the gradient operator. In this paper, bold notations such as $\bm{x}$ shall denote vector-based notations, and blackboard faces such as ${\mathbb{X}}$ sets of (tuples of) real numbers or natural integers of ${\mathbb{R}}$ or ${\mathbb{N}}$ respectively. $D_\varphi(\bm{x} || \bm{y})$ is the difference between the value of $\varphi$ at $\bm{x}$ and the value at $\bm{x}$ of the hyperplane tangent to $\varphi$ in $\bm{y}$. Bregman divergences encode a natural notion of distortion, as shown by Theorem \ref{th0} below. Its proof is a slight variation of Theorem 4 in \cite{bgwOT} (see also \cite{bmdgCWj}).
\begin{theorem}\label{th0}
Let $F : {\mathbb{R}}^d \times {\mathbb{R}}^d \rightarrow {\mathbb{R}}$ be a function that satisfies the following three axioms ($\forall \bm{x}, \bm{y} \in {\mathbb{R}}^d$):
\begin{enumerate}
\item non-negativity: $F(\bm{x}, \bm{y}) \geq 0$; 
\item identity of indiscernibles: $F(\bm{x}, \bm{y}) = 0$ if and only if $\bm{x} = \bm{y}$;
\item the expectation is the lowest distortion's predictor: for any random variable $\bm{X}$ whose distribution ${\mathcal{D}}$ has support ${\mathbb{R}}^d$,
\begin{eqnarray}
\arg_{\bm{y} \in {\mathbb{R}}^d} \min \es_{{\mathcal{D}}} F(\bm{X}, \bm{y}) & = & \es_{{\mathcal{D}}} \bm{X} (\defeq \mu_{\bm{X}}) \label{defmin} \:\:,
\end{eqnarray}
where $\es_{{\mathcal{D}}}$ denotes the mathematical expectation.
\end{enumerate}
Then $F(\bm{x}, \bm{y}) = D_\varphi(\bm{x}|| \bm{y})$ for some strictly convex and differentiable $\varphi: {\mathbb{R}}^d \rightarrow {\mathbb{R}}$.
\end{theorem}
(proof: see the Appendix) It is easy to check that any Bregman divergence satisfies [1], [2] and [3] \cite{bmdgCWj}, and so Theorem \ref{th0} provides a complete characterization of Bregman divergences, in the same way as conditions [1] and [2], completed with symmetry and subadditivity, would axiomatize a metric. This positions Bregman divergences with respect to numerous metric-related notions, and gives the importance of their main difference, eq. (\ref{defmin}). Eq. (\ref{defmin}) is fundamental because it says that the (arithmetic) expectation is the lowest distortion parameter for a population, regardless of the distortion. Actually, eq. (\ref{defmin}) says much more: the expectation is \textit{maximum likelihood estimator} of data for a large set of distributions called the \textit{exponential families}. These families contain some of the most popular distributions, such as Bernoulli, multinomial, beta, gamma, normal, Rayleigh, Laplacian, Poisson \cite{bmdgCWj,nbnOB}. A remarkable property is that any member satisfies the following identity \cite{bmdgCWj}:
\begin{eqnarray}
\log \pr[\bm{x} | \bm{\theta}, \varphi] & = & -D_\varphi(\bm{x} || \bm{\mu}_{\bm{\theta}}) + \log b_\varphi(\bm{x}) \:\:. \label{defexp}
\end{eqnarray}
$\bm{\theta}$ defines the so-called natural parameters of the distribution, and $b_\varphi(.)$ is a normalization function. It follows from (\ref{defexp}) and (\ref{defmin}) that the maximum likelihood estimator of data is the expectation parameter $\bm{\mu}_{\bm{\theta}}$.

\begin{table}
\centering
\begin{tabular}{lccl}\hline \hline
$\mathrm{dom}(\varphi)$ & $\phi(x)$  & $D_\varphi(\bm{x}|| \bm{y})$ & Divergence name\\ \hline
${\mathbb{R}}^d$ & $x^2$ & $\sum_{i=1}^d{(x_i - y_i)^2}$ & Squared Euclidean norm \\ \hline
${\mathbb{R}}_+^d$ & $x \log x - x$ & $\sum_{i=1}^d{x_i \log\frac{x_i}{y_i} - x_i + y_i}$ & Kullback-Leibler div. \\
${{\mathbb{P}}}_d$ & id. & $\sum_{i=1}^d{x_i \log\frac{x_i}{y_i}}$ & Entropy \\ \hline
${\mathbb{R}}_+^d$ & $-\log x$ & $\sum_{i=1}^d{\frac{x_i}{y_i} - \log\frac{x_i}{y_i} - 1}$ & Itakura-Saito div. \\ \hline
$[0,1]$ & $\begin{array}{c}
x\log x\\
+ (1-x) \log(1-x)
\end{array}$ & $
\begin{array}{c}
x\log\frac{x}{y}\\
+ (1-x) \log\frac{1-x}{1-y}
\end{array}$ & Logistic loss \\\hline \hline
\end{tabular}
\caption{Correspondence between various generators and their Bregman divergences. ${{\mathbb{P}}}_d$ is the $d$-dimensional probability simplex. The generator of the Bregman divergence is defined by $\varphi(\bm{x}) \defeq \sum_{i=1}^{d}{\phi(x_i)}$ (see text).\label{t-exe}}
\end{table}
Some Bregman divergences have become cornerstones of various fields of mathematics and computer science, as shown in Table \ref{t-exe}. All of them are \textit{separable} Bregman divergences \cite{dsGN}, as they can be characterized using a strictly convex function $\phi: {\mathbb{I}} \subseteq {\mathbb{R}} \rightarrow {\mathbb{R}}$, the generator of the Bregman divergence being just:
\begin{eqnarray}
\varphi(\bm{x}) & \defeq & \sum_{i} {\phi(x_i)} \:\:. \label{defsep}
\end{eqnarray} 

It might seem that (\ref{defmin}) unveils a strong assymetry between the two parameters of a Bregman divergence, all the more as that Bregman divergences are not symmetric in almost all cases \cite{nbnOB}. This distinction becomes more superficial --- but crucial for our purpose --- as \textit{Legendre duality} enters the analysis. Any Bregman divergence is indeed equal to a Bregman divergence over \textit{swapped} parameters in the generator's gradient space. To make it formal, the generator $\varphi$ of a Bregman divergence admits a convex conjugate $\varphi^\star : {\mathbb{R}}^d \rightarrow {\mathbb{R}}$ given by \cite{rCA}:
\begin{eqnarray}
\varphi^\star(\bm{y}) & \defeq & \sup_{\bm{x} \in {\mathbb{X}}} \{ \langle \bm{x}, \bm{y} \rangle - \varphi(\bm{x})\} \label{defstar}\\
 & = & \langle \bm{y}, \bm{\nabla}^{-1}_{\varphi} (\bm{y}) \rangle - \varphi\left( \bm{\nabla}^{-1}_{\varphi} (\bm{y}) \right) \label{deriv}\:\:,
\end{eqnarray}
where $\bm{\nabla}^{-1}_\varphi$, the inverse gradient, is well-defined because of the strict convexity of $\varphi$. The following Theorem, whose proof follows from plugging (\ref{deriv}) in (\ref{bdir}), states the dual symmetry of Bregman divergences.
\begin{theorem}
$D_\varphi(\bm{x} || \bm{y}) = D_{\varphi^\star}(\bm{\nabla}_\varphi(\bm{y}) || \bm{\nabla}_\varphi(\bm{x}))$.
\end{theorem}
It follows from (\ref{defmin}) and the strict convexity of $\varphi$ that the minimizer of the expected dual divergence $D_{\varphi^\star}(.||.)$ can be expressed in $\mathrm{int}(\mathrm{dom}(\varphi))$ as:
\begin{eqnarray}
\bm{\mu}_{\varphi} & \defeq & \bm{\nabla}^{-1}_{\varphi} (\es_{{\mathcal{D}}} \bm{\nabla}_\varphi(\bm{X})) \:\:.\label{bregmean}
\end{eqnarray}
The set spanned by (\ref{bregmean}), which includes the arithmetic average (take $\phi \defeq x^2/2$ in (\ref{defsep})), is close to the set of $f$-means \cite{hlpI}, a set whose studies date back to the early thirties, by Kolmogorov and Nagumo.

To summarize the conceptual justifications for the use of \textit{aggregators} having shape (\ref{bregmean}), three main motivations could justify their use: first, they are all optimal distortion estimators --- and the only ones to be optimal -- in the sense of Theorem \ref{th0}; second, they encode maximum likelihood estimators for a majority of popular distributions; third, they encode geodesic-like curves in the geometry of the information space \cite{nbnOB}. For all these reasons, they can be considered the best information aggregators for the data they summarize (data which could be prices, consumptions, labors, wages, etc. in the economic world).

Hereafter, we consider averages (\ref{bregmean}) with finite support of size $m>0$, and replace (\ref{defmin}) by the more general search for $\arg_{\bm{y} \in {\mathbb{R}}^d} \min \sum_{i=1}^{m} {\gamma_i F(\bm{x}_i, \bm{y})}$, with $\gamma_i > 0, \forall i=1, 2, ..., m$. A rapid glimpse at (\ref{defmin}) reveals that the solution is $\sum_{i=1}^{m} {\gamma_i \bm{x}_i}$, and so the extension of (\ref{bregmean}) to the minimizer of a general weighted sum of Bregman divergences now takes the more general form:
\begin{eqnarray}
\bm{\mu}_{\varphi} & \defeq & \Gamma \bm{\nabla}^{-1}_{\varphi} \left(\frac{1}{\Gamma} \sum_{i=1}^{m} {\gamma_i \bm{\nabla}_\varphi(\bm{x}_i)}\right) \:\:,\label{bregmeangen}
\end{eqnarray}
with $\Gamma \defeq \sum_{i=1}^{m} {\gamma_i}$. Because of (\ref{defmin}), any $\bm{\mu}_\varphi$ as in (\ref{bregmeangen}) is called a low-distortion aggregator (LDA). For economic and mathematical reasons, averages having the form (\ref{bregmeangen}) with a concave or convex regime are particularly interesting. The following Theorem allows to catch the picture of where concavity and convexity lie: the symmetric dual average of some average (\ref{bregmeangen}) enjoys the symmetric regime. If one is concave, the other is convex and \textit{vice versa}. 
\begin{theorem}\label{th1}
$\mu_\varphi$ is concave if and only if $\mu_{\varphi^\star}$ is convex.
\end{theorem}
(proof: see the Appendix). To finish up with information, we state the last result that shall be useful in the sequel.
\begin{theorem}\label{thagh}
Let $\mu_{\varphi}$ a concave (resp. convex) average that follows (\ref{bregmean}). Then it is upperbounded (resp. lowerbounded) by the sum: $\bm{s} = \sum_{i=1}^{m} {\gamma_i \bm{x}_i}$. Furthermore, $\bm{\nabla}_\varphi$ is concave (resp. convex), and $\bm{\nabla}^{-1}_\varphi$ is convex (resp. concave).
\end{theorem}
(proof: see the Appendix). 

\section{Economic Aggregators}

Because a LDA $\mu_\varphi$ does not change by adding a constant term to its generator $\varphi$, it should be kept in mind that generators shall be given up to any such constant. Furthermore, our analysis takes place for separable generators, that meet (\ref{defsep}). This eases readability while encompassing most economic settings. For such reasons, it is also convenient to assume that $\mathrm{dom}\phi \subseteq {\mathbb{R}}_+$, and introduce the following notation for any relevant $k \in {\mathbb{N}}_*$:
\begin{eqnarray}
\phi^{[k]}(x) & \defeq & \frac{\mathrm{d}^k \phi(x)}{\mathrm{d} x^k}\:\:.
\end{eqnarray}

\subsection{Optimality of Economic Aggregators}

Let $x_\star$ denote an aggregator for values $x_1, x_2, ..., x_m$ ($m\in {\mathbb{N}}_*$). One of the most common economic aggregators is the CES function \cite{acmsCL}:
\begin{eqnarray}
x_\star & \defeq & \left(\sum_{i=1}^{m} \beta_i x_i^\frac{\sigma-1}{\sigma} \right)^{\frac{\sigma}{\sigma - 1}}\:\:. \label{defgences}
\end{eqnarray}
The formulation in $\sigma$ is not the simplest but it is intentional, as it depicts the constant elasticity of substitution inside values aggregated \cite{bkMC}. Here, $\beta_i > 0$ is the weight of aggregated value $x_i$. Further constraints of economic relevance are generally imposed on $\sigma$ depending on the setting in which (\ref{defgences}) is applied \cite{bkMC}; in order to remain as general as possible, we consider the unrestricted setting for which $\sigma \in {\mathbb{R}}_* \backslash \{1\}$. We now show that a CES is a LDA.
\begin{lemma}\label{lemcstar}
Any CES $x_\star$ as defined in (\ref{defgences}) is a LDA for the generator 
\begin{eqnarray}
\phi_{\mathrm{\textsc{ces}}}(x) & \defeq & a x^{2 - \frac{1}{\sigma}}\:\:, \label{propnces}
\end{eqnarray} 
with $a\in {\mathbb{R}}_*$ any constant for which (\ref{propnces}) is convex, and $\gamma_i \defeq \beta_i \Beta^{\frac{1}{\sigma-1}}, \forall i = 1, 2, ..., m$. Here, $\Beta \defeq \sum_{i=1}^{m} {\beta_i}$.
\end{lemma}
(proof: see the Appendix). Aggregators are sometimes tied up via important economic equalities. One example relates prices and consumptions. Let $m$ denote the number of goods, and the consumption function of good $i$ is noted $c_i$. The price of good $i$ is $p_i$. Two aggregators for consumptions and prices, respectively $c_\star$ and $p_\star$ are devised so as to satisfy:
\begin{eqnarray}
\sum_{i=1}^{m} {c_i p_i} = p_\star c_\star\:\:. \label{fixp}
\end{eqnarray}
Further economic assumptions can be made, such as the concavity of $c_\star$, which indicates the preference for diversity \cite{dsMC}. The popular choice for $c_\star$ is a CES function (\ref{defgences}) \cite{acmsCL}. Notice that the weights in the LDA ($\gamma_i$) are different from the weights in the CES ($\beta_i$). Modulo a simple normalization of the CES, they remain equal. If we multiply $c_\star$ by $\Beta^{1/(1-\sigma)}$, the normalized CES obtained is such that $\gamma_i = \beta_i$. Furthermore, this normalization, for which $\Beta^{1/(1-\sigma)} = m^{1/(1-\sigma)}$ when all $\beta_i = 1$, is one which turns out to play a key role in economic models \cite{bkMC}.

The price aggregator, $p_\star$, can be found by inspecting (\ref{fixp}) after remarking that partial derivatives on the left and right-hand side must also coincide. After a standard derivation using (\ref{defgences}) for $c_\star$, we obtain that the price index has the form:
\begin{eqnarray}
p_\star & = & \left( \sum_{i=1}^{m} {\beta_i^\sigma p_i^{1-\sigma}} \right)^{\frac{1}{1-\sigma}} \:\:.\label{defdualp}
\end{eqnarray}
$p_\star$ has also the general CES form of (\ref{defgences}); for completeness, we characterize below its LDA (proof similar to Lemma \ref{lemcstar}).
\begin{lemma}\label{lempstar}
Let $\delta_i \defeq \beta_i^\sigma$, and $\Delta \defeq \sum_{i=1}^{m} {\delta_i}$. The price index in (\ref{defdualp}) is a LDA for the generator:
\begin{eqnarray}
\phi_{p}(x) & \defeq & b x ^{2-\sigma}\:\:, \label{propprix}
\end{eqnarray}
with $b\in {\mathbb{R}}_*$ any constant for which (\ref{propprix}) is convex, and $\gamma_i \defeq \delta_i \Delta^{\frac{\sigma}{1 - \sigma}}, \forall i = 1, 2, ..., m$.
\end{lemma}
Modulo the normalization of the CES for $c_\star$, and the choice $\beta_i = 1$, (\ref{defdualp}) would return to the conventional choice in which $\beta_i^\sigma \rightarrow 1/m$. It is quite a remarkable fact that $p_\star$ and $c_\star$ are LDA under the sole assumptions of (\ref{fixp}) and $c_\star$ is a CES. Such a property also holds for labor and wages. Suppose we have $n$ consumer-workers, each of which selling a particular labor type; let $w_j$ be the wage for labor-type $j$ and $n_j$ the demand for labor-type $j$, for $j=1, 2, ..., n$. Then there exists an aggregate labor-demand index $n_\star$, and a wage index $w_\star$, such that \cite{bkMC}:
\begin{eqnarray}
\sum_{j=1}^{n} {w_i n_i} & = & w_\star n_\star \:\:. \label{fixn}
\end{eqnarray}
The CES form for $w_\star$ \cite{bkMC} implies both the LDA property for $w_\star$ and $n_\star$ (Lemmata \ref{lemcstar} and \ref{lempstar}). To summarize, popular aggregators for consumptions, prices, labor and wages are all LDAs, which means that they are all optimal from the information theory standpoint. Before drilling down further into the properties that yield relationships like (\ref{fixp}) or (\ref{fixn}), let us give a brief panorama of which Bregman divergences are involved so far.

The Bregman divergence of a CES (\ref{defgences}) is:
\begin{eqnarray}
D_{\phi_{\mathrm{\textsc{ces}}}}(\bm{x} || \bm{z}) & = & a\sum_{i=1}^{m}{\left\lbrace x_i^{2 - \frac{1}{\sigma}} - \left( 2 - \frac{1}{\sigma} \right)x_i z_i^{1-\frac{1}{\sigma}} + \left(1-\frac{1}{\sigma}\right)z_i^{2-\frac{1}{\sigma}} \right\rbrace} \:\:.
\end{eqnarray}
Since any CES is a LDA, it follows that Cobb-Douglas and Leontief functions are limit LDAs, respectively when $\sigma \rightarrow 1$ and $\sigma \rightarrow 0^+$. While Leontief function, $x_\star \defeq \min_i \{\beta_i x_i\}$, does not admit a generator (it is not differentiable), Cobb-Douglas,
\begin{eqnarray}
x_\star & \defeq & \prod_{i=1}^{m} {x_i^{\beta_i}} \:\:, \label{defcd}
\end{eqnarray}
admits one, which is:
\begin{eqnarray}
\phi_{\mathrm{\textsc{cd}}}(x) & \defeq & b (x\log x - x) \label{cobb}\\
 &  & = b \phi_{\mathrm{\textsc{kl}}} (x) \nonumber
\end{eqnarray} 
(see Table \ref{t-exe}; $b \in {\mathbb{R}}_{+,*}$ is any constant). If we look at the price index in (\ref{defdualp}), we get the following result.
\begin{lemma}
Fix $b=1/((2-\sigma)(1-\sigma))$ in (\ref{propprix}), assuming $\sigma \neq 1$, and let $\phi_{\mathrm{IS}} \defeq -\log x$ (see Table \ref{t-exe}). Then 
\begin{eqnarray}
\lim_{\sigma \rightarrow 2} D_{\varphi_{p}}(\bm{x} || \bm{z}) & = & D_{\varphi_{\mathrm{IS}}} (\bm{x} || \bm{z})\:\:. \label{limit1}
\end{eqnarray}
\end{lemma}
This result is easily proven once we remark that $x^{k} \approx 1 + k\log x + o(k)$. The right-hand side of (\ref{limit1}) is Itakura-Saito divergence (Table \ref{t-exe}). Together with the fact that the limit divergence for $c_\star$ is Kullback-Leibler divergence when $\sigma \rightarrow 1$, we get the generators for two popular divergences of signal processing and statistics \cite{nbnOB}. A well-known similar result holds for a particular subset of Bregman divergences, Amari $\alpha$-divergences, for which \cite{anIG}:
\begin{eqnarray}
\phi_{\mathrm{\textsc{a}}} (x) & \defeq & 4\left(x - x^\frac{1+\alpha}{2}\right)/(1-\alpha^2), \alpha \in [-1,1] \:\:.
\end{eqnarray} 
Taking limits of the generator when $\alpha$ reaches the interval bounds yields Itakura-Saito and Kullback-Leibler divergences.

\subsection{Completeness of Economic Aggregators}

In this section, we consider some relevant economic assumptions about aggregators, and show that any LDA that would meet such assumptions would necessarily belong to a particular subclass of LDAs. This subclass is called ``complete'' for the assumption at hand.

The first assumption we consider is about any two \textit{dual} aggregators $x_\star$ (for $x_1, x_2, ..., x_m$) and $z_\star$ (for $z_1, z_2, ..., z_m$) that would meet the following abstraction of (\ref{fixp}) and (\ref{fixn}):
\begin{eqnarray}
\sum_{i=1}^{m} {x_i z_i} & = & x_\star z_\star \:\:. \label{fixg}
\end{eqnarray}
We show that CES turns out to be complete for dual aggregators, as the LDA assumption for any of the two implies that \textit{both} are CES. We state it more formally below.
\begin{theorem}\label{fundth}
Suppose that at least one of $x_\star$ and $z_\star$ that satisfies (\ref{fixg}) is a LDA. Then both $x_\star$ and $z_\star$ are CES. Furthermore, they are linked through the identity $\phi^{[2]}_z = d \left(\phi^{[2]}_x\right)^{-1}$ for some $d \in {\mathbb{R}}_*$.
\end{theorem}
(proof: see the Appendix). CES turns out to be complete from another standpoint: elasticities. Consider some LDA $x_\star$; its elasticity with respect to $x_i$ ($i=1, 2, ..., m$) is defined as:
\begin{eqnarray}
{{\mathrm{\textsc{e}}}}_{x_\star}^{x_i} & \defeq & \left(\frac{\mathrm{d} x_\star}{x_\star}\right)/ \left(\frac{\mathrm{d} x_i}{x_i}\right)\:\:. \label{defelast}
\end{eqnarray}
Consider the economic assumption that all elasticities sum to one. We show that CES is complete for this assumption. 
\begin{theorem}\label{constelast}
Let $x_\star(x_1, x_2, ..., x_m)$ be any LDA. Then $\sum_{i=1}^{m} {{{\mathrm{\textsc{e}}}}_{x_\star}^{x_i}} = 1$ if and only if $x_\star$ is a CES.
\end{theorem}
(proof: see the Appendix). We now switch to another important economic quantity, the substitution elasticity of $x_i$ for $x_j$ in $x_\star$, ${\mathrm{\textsc{e}}}_{x_\star}^{x_i \rightarrow x_j}$, defined by:
\begin{eqnarray}
{\mathrm{\textsc{e}}}_{x_\star}^{x_i \rightarrow x_j} & \defeq & \left( \frac{\mathrm{d}(x_j / x_i)}{x_j/x_i} \right) / \left( \frac{\mathrm{d}{\mathrm{\textsc{s}}}_{x_\star}^{x_i \rightarrow x_j}}{{\mathrm{\textsc{s}}}_{x_\star}^{x_i \rightarrow x_j}} \right)\:\:, \label{defelsub}
\end{eqnarray}
where 
\begin{eqnarray}
{\mathrm{\textsc{s}}}_{x_\star}^{x_i \rightarrow x_j} & \defeq & \left( \frac{\partial x_\star}{\partial x_i} \right) / \left( \frac{\partial x_\star}{\partial x_j} \right)
\end{eqnarray}
is the marginal rate of substitution of $x_i$ for $x_j$. Another economic assumption commonly encountered is the fact that ${\mathrm{\textsc{e}}}_{x_\star}^{x_i \rightarrow x_j}$ is assumed to be unit. We show that the complete LDA subclass for this assumption is, this time, Cobb-Douglas. 
\begin{theorem}\label{constsubs}
Let $x_\star(x_1, x_2, ..., x_m)$ be any LDA. Then, there exists indices $1\leq i,j\leq m$ such that ${\mathrm{\textsc{e}}}_{x_\star}^{x_i \rightarrow x_j} = 1$ if and only if $x_\star$ is a Cobb-Douglas.
\end{theorem}
(proof: see the Appendix). It is interesting to notice that the LDA assumption competes with the homogeneity assumptions about $x_\star$ that are required to come up with the same result (i.e. without making the LDA assumption). The fact that we are able to alleviate the economic setting (homogeneity ties up $x_\star$ with assumptions on returns to scale) while ending up with the same aggregator makes information a very valuable companion to introduce the true nature of popular economic aggregators. One question which remains is however what would imply the homogeneity assumption \textit{alone} in a LDA setting. We define $x_\star$ to be homogeneous of degree $a \in {\mathbb{R}}_*$ if and only if:
\begin{eqnarray}
x_\star(\lambda x_1, \lambda x_2, ..., \lambda x_m) & = & \lambda^a x_\star(x_1, x_2, ..., x_m) \:\:, \label{defhom}
\end{eqnarray}
for every $\lambda \in {\mathbb{R}}_+$. We show that the complete LDA subclass for this assumption varies depending on the values of $a$. Without losing too much generality, the Theorem assumes that $\phi^{[2]}_x$ is differentiable.
\begin{theorem}\label{consthom}
Let $x_\star(x_1, x_2, ..., x_m)$ be any LDA, and $a\in {\mathbb{R}}_*$. Then:
\begin{itemize}
\item $x_\star$ is homogeneous of degree $a \neq 1$ if and only if it is a Cobb-Douglas;
\item $x_\star$ is homogeneous of degree $a = 1$ if and only if it is a Cobb-Douglas or a CES.
\end{itemize}
\end{theorem}
(proof: see the Appendix). 

\section{Discussion}

\begin{table*}[t]
\begin{center}
\begin{tabular}{r|c|ccccc} \hline\hline
 & Optimality & \multicolumn{5}{|c}{Completeness} \\ 
 &  (LDA)   & Th. \ref{fundth} & Th. \ref{constelast} & Th. \ref{constsubs} & Th. \ref{consthom}& Th. \ref{consthom}\\ 
 &   &  & &  & ($a\neq 1$) & ($a=1$)\\ \hline
CES          & Y & Y & Y & N & N & Y \\ 
Cobb-Douglas & Y & N & N & Y & Y & Y \\ 
Leontief     & L & L & L & N & N & L \\ 
MST          & N & N & N & N & N & N \\ \hline \hline
\end{tabular}
\end{center}
\caption{Summary of our results on four families of aggregators: CES, Cobb-Douglas, Leontief and MST, with respect to the assumptions made in Theorems \ref{fundth}, \ref{constelast}, \ref{constsubs} and \ref{consthom} (see text for details).\label{t-sum}}
\end{table*}

\paragraph{Families of economic aggregators} Table \ref{t-sum} summarizes the results obtained on three families of aggregators: CES, Cobb-Douglas and Leontief. For each of them we give the indication of whether they are LDAs (Y/N), whether they can be in the limit (L), and whether they become complete for the assumptions made in Theorems \ref{fundth}, \ref{constelast}, \ref{constsubs} and \ref{consthom} (Y / N / L). Remark that the Table makes a clear distinction between all these three families of aggregators. There exists various other aggregators in economic works; for obvious space reasons, we have chosen to focus on the most popular, and it turns out that all have strong relationships with LDAs, either directly, or at the limit. In order to cover the possible relationships between aggregators and LDAs, let us take a last example, of a general class of aggregators that we call Mitscherlich-Spillman-von Th{\"u}nen (MST) aggregators \cite{mDG,sTL,tDI}, a family in which the global form of aggregator $x_\star$ reduces directly or after a variable change to ($\theta \in \{-1,+1\}$):
\begin{eqnarray}
x_\star & \defeq & \prod_{i=1}^{m} {(1 - \exp(\theta \gamma_i x_i))} \:\:. \label{mst}
\end{eqnarray}\label{mstlda}
Such aggregators date back to the XIX$^{th}$ century, and so they have preceded those we have been focusing on so far. What we can show is that, contrasting with their successors, MST aggregators are \textit{not} LDAs.
\begin{lemma}
MST aggregators are not LDAs.
\end{lemma}
\begin{figure*}[t]
\begin{center}
\begin{tabular}{c}
\epsfig{file=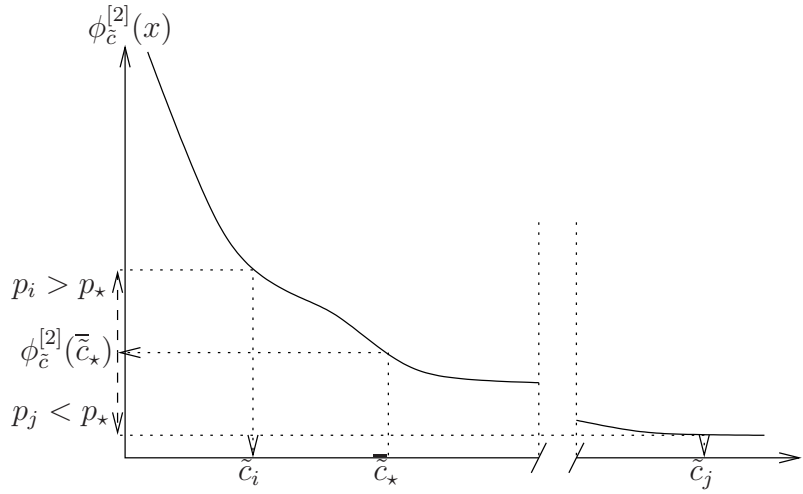}
\end{tabular}
\end{center}
\caption{Prices vs consumptions for a concave consumption LDA in (\ref{genpctilde}): regardless of the LDA, for any good $i$, if its price is larger (resp. smaller) than the price index, its consumption cannot be larger (resp. smaller) than the normalized consumption index (see text for details).\label{f-prices}}
\end{figure*}
(proof: see the Appendix). 

\paragraph{Aggregators and economic constraints} Modulo changes of variables, Theorems (\ref{fundth}) - (\ref{consthom}) could be alleviated from the constraint of the LDA choice under their respective economic assumptions. Consider for example (\ref{fixg}), in which we would like to plug any LDA. To be concrete, let us stick to prices and consumptions in (\ref{fixp}). Consider the generator $\phi_{\tilde{c}}$ of some strictly concave LDA $\tilde{c}_\star$ that aggregates its consumptions $\tilde{c}_i$ for $i=1, 2, ..., m$. Let us say that strict concavity is chosen because usual consumption indexes are concave, to indicate the consumer's preference for diversity \cite{dsMC}. Consider the change of variables that involves some (concave) CES: 
\begin{eqnarray}
c_i & \defeq & \left(\phi^{[1]}_{\mathrm{\textsc{ces}}}\right)^{-1} \left( \phi^{[1]}_{\tilde{c}} (\tilde{c}_i) \right)\:\:.\label{def14}
\end{eqnarray}
If we consider that consumptions $c_i$ are the actual observed consumptions (the $\tilde{c}_i$'s being ``hidden'', non-observed consumptions), the aggregator $c_\star$ for those lifted consumptions is a CES which may be plugged in (\ref{fixp}). Reconstructing $\tilde{c}_\star$ is immediate as we have:
\begin{eqnarray}
\tilde{c}_\star & = & \left(\phi^{[1]}_{\tilde{c}}\right)^{-1} \left(\phi^{[1]}_{\mathrm{\textsc{ces}}}(c_\star) \right) \:\:.\label{def15}
\end{eqnarray}
In order not to laden the discussion, let us consider that all weights $\beta_i = 1/m$, and that we keep the CES form for $p_\star$ ($p_i = \tilde{p}_i$). What interests us here is not exactly the consumption index ($c_\star$), but the \textit{normalized} index, $\overline{c}_\star \defeq c_\star / m$, which is really homogeneous to the consumption of a single good. We also define the same index $\overline{\tilde{c}}_\star$ for $\tilde{c}_\star$, and assume the same relationship as (\ref{def15}) for these two indices. Differentiating (\ref{fixp}) in $c_i$ yields:
\begin{eqnarray}
\frac{p_i}{p_\star} & = & \frac{\phi^{[2]}_{\mathrm{\textsc{ces}}}(c_i)}{\phi^{[2]}_{\mathrm{\textsc{ces}}}(\overline{c}_\star)} \:\:, \label{relpgen}
\end{eqnarray}
from which we obtain using (\ref{def14}) and (\ref{def15}):
\begin{eqnarray}
\frac{p_i}{p_\star} & = & \frac{\phi^{[2]}_{\tilde{c}}(\tilde{c}_i)}{\phi^{[2]}_{\tilde{c}}(\overline{\tilde{c}}_\star)}\:\:. \label{genpctilde}
\end{eqnarray}
Eq. (\ref{genpctilde}) is interesting because it displays a remarkably stable behavior that holds for \textit{any} concave LDA. Because of Theorem \ref{thagh} and the convexity of $\phi$, $\phi^{[2]}_{\tilde{c}}$ is monotonic decreasing and strictly positive. Thus, it converges towards some non negative value.
Figure \ref{f-prices} gives an overview of how prices and consumptions always behave. There are two conclusions to draw from the figure. The first is a sanity check, as larger prices mean lower consumptions, a conclusion that follows from dividing (\ref{genpctilde}) for distinct goods $i$ and $j$. Indeed, if a good $i$ has price $p_i > p_j$, then we shall have $\tilde{c}_i \leq \tilde{c}_j$ (notice, from (\ref{def14}), that it is equivalent to saying $c_i \leq c_j$). We also remark that prices that are larger (resp. smaller) than the price index mean consumptions that are smaller (resp. larger) than the normalized consumption index (see Figure \ref{f-prices}). The second is the dampening effect of prices on consumptions: a small difference on prices may incur a very large difference on consumptions if those prices are already small, and it can make almost no difference on consumptions if prices are high.

\paragraph{Global behaviors of LDAs as economic aggregators} Even without a change of variables, LDAs sometimes display economic regimes with extremely close behaviors, as witnessed by the marginal rate of substitution of $x_i$ for $x_j$. Indeed, whenever $x_\star$ is a LDA, we have:
\begin{eqnarray}
{\mathrm{\textsc{s}}}_{x_\star}^{x_i \rightarrow x_j} & = & \frac{\gamma_i \phi^{[2]}_x(x_i)}{\gamma_j \phi^{[2]}_x(x_j)}\:\:.
\end{eqnarray}
\begin{figure*}[t]
\begin{center}
\begin{tabular}{c}
\epsfig{file=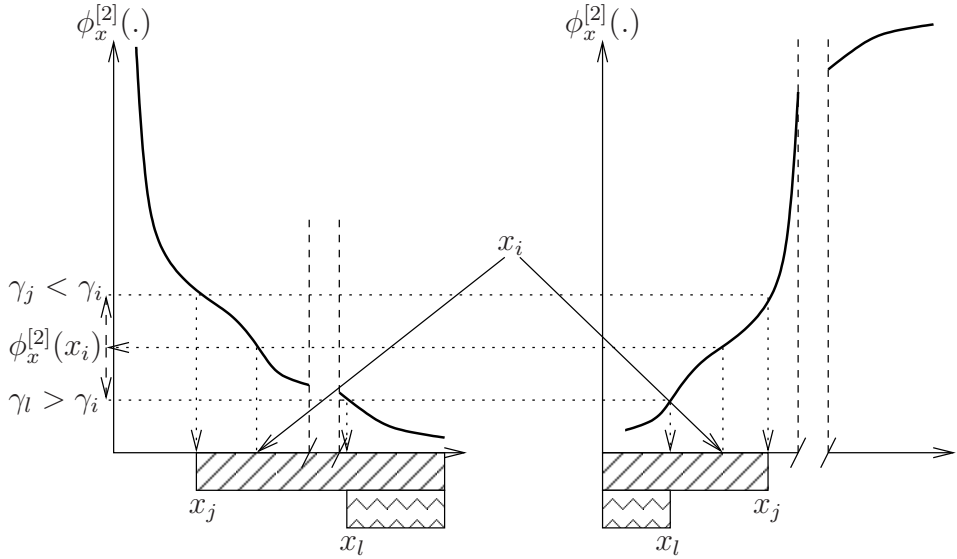}
\end{tabular}
\end{center}
\caption{Intervals for which the marginal rate of substitution of $x_i$ for $x_j$ (or $x_l$) exceeds unit when $x_\star$ is a concave (left) or convex (right) LDA.\label{f-mrs}}
\end{figure*}
Figure \ref{f-mrs} displays the general behaviors of ${\mathrm{\textsc{s}}}_{x_\star}^{x_i \rightarrow x_.}$ as a function of the concavity or convexity of $x_\star$. In the convex case, Theorem \ref{thagh} and the convexity of $\phi$ bring that $\phi^{[2]}_{x}$ is monotonic increasing and strictly positive, hence the schema depicted in Figure \ref{f-mrs} (right). The dashed rectangles depict the intervals for which this marginal rate of substitution would be greater than 1, that is, locations where we would be willing to trade more than one unit of $x_.$ to obtain one unit of $x_i$. The behavior is remarkably linked with the global regime of $x_\star$: when it is concave, Figure \ref{f-mrs} (left) clearly displays a preference for diversity, while when it is convex, Figure \ref{f-mrs} (right) shows the symmetric trend, an aversion for diversity.

\begin{figure*}[t]
\begin{center}
\begin{tabular}{c}
\epsfig{file=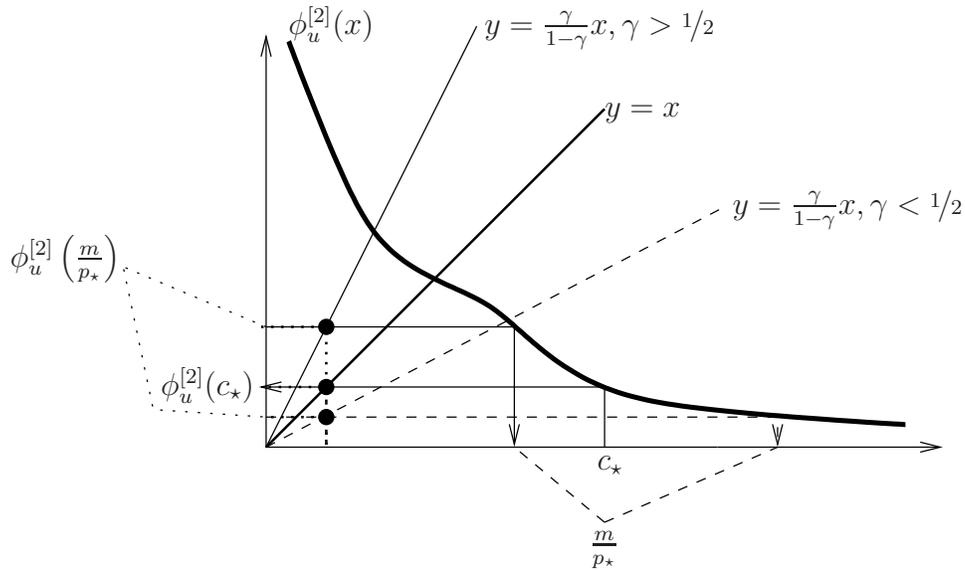}
\end{tabular}
\end{center}
\caption{Depiction of the relationships between consumption and money for the consumer's program (\ref{fixu}) (when $\gamma>\nicefrac{1}{2}$ and $\gamma < \nicefrac{1}{2}$; see text for details).\label{f-util}}
\end{figure*}

\paragraph{Aggregators of aggregators and economic programs} LDAs may incorporate heterogeneous quantities and even LDAs as well, as it is common for economic aggregators to integrate other economic aggregators. Examples show how the whole aggregate may behave, and how global regimes underlined above for ``baseline'' aggregators also emerge in a simple manner for whole aggregates as well. Consider the determination of the global consumption index $c_\star$ and money expenses $m$ (hereafter, $m$ does not refer anymore to the number of aggregated values) of a consumer, based on his/her whole budget $r$ and price indices $p_\star$ (we do not make any assumption on the form of $c_\star$ and $p_\star$). The consumer solves the maximization of a utility aggregator $u_\star$:
\begin{eqnarray}
\max_{c_\star, m} u_\star & \mbox{ s.t. } & p_\star c_\star + m = r\:\:. \label{fixu}
\end{eqnarray}
Let us investigate the general solution of (\ref{fixu}), under the sole assumption that $u_\star$ is some concave LDA that mixes consumption and money via some Bernoulli distribution $B(\gamma)$ for coefficients $\gamma_i, i=1, 2$ which leverages the importance of consumption and money in $u_\star$:
\begin{eqnarray}
u_\star & = & \nabla^{-1}_{\phi_u}\left( \gamma \nabla_{\phi_u}\left(c_\star\right) + (1-\gamma) \nabla_{\phi_u}\left(m/p_\star\right) \right) \:\:. \label{defutil}
\end{eqnarray}
The following Theorem states the fundamental relationships that may be used to determine $c_\star$ and $m$, for any concave LDA.
\begin{theorem}\label{thutil}
The optimal values for $c_\star$ and $m$ in (\ref{fixu}) satisfy:
\begin{eqnarray}
\phi^{[2]}_u \left(c_\star\right) & = & \frac{1 - \gamma}{\gamma} \phi^{[2]}_u\left(\frac{r}{p_\star} - c_\star\right) \:\:, \label{eqcs}\\
\phi^{[2]}_u\left(\frac{m}{p_\star}\right) & = & \frac{\gamma}{1 - \gamma} \phi^{[2]}_u\left(\frac{r}{p_\star} - \frac{m}{p_\star}\right)  \:\:. \label{eqm}
\end{eqnarray}
\end{theorem}
(proof: see the Appendix). Because $\phi^{[2]}_u$ is monotonous decreasing and strictly positive, solving (\ref{eqcs}) and (\ref{eqm}) can be done via a simple dichotomic search in the general case where $\phi^{[2]}(x)$ would be strictly monotonous (when strict monotonicity is not ensured, we may logically end up with an interval of values). The proof of Theorem \ref{thutil} reveals an interesting relationship between consumption and money, namely:
\begin{eqnarray}
\phi^{[2]}_u(c_\star) & = & \frac{1-\gamma}{\gamma} \phi^{[2]}_u \left(\frac{m}{p_\star} \right) \:\:. \label{relcm}
\end{eqnarray}
Figure \ref{f-util} displays this relationship, and more precisely where $m/p_\star$ is located with respect to $c_\star$, depending on $\gamma$. Remark that when $\gamma > \nicefrac{1}{2}$, which marks the predominance of consumption over money in the consumers' utility function $u_\star$, Figure \ref{f-util} shows that $c_\star$ indeed represents more than money in the whole budget, as we recall that $c_\star + (m/p_\star) = r/p_\star$. The symmetric situation holds when $\gamma < \nicefrac{1}{2}$. 

\section{Conclusion}

The aim of this paper was to demonstrate that the choice of various economic aggregators, that have mainly been originally built on empirical economic grounds, meet optimality from the information standpoint, and most notably, remain the \textit{only} optimal aggregators as various standard economic assumptions are considered. In these settings, they are the only ways one could summarize at best numerous individual variables in aggregate indexes, designed to represent their collective economic behavior at the global level. We have also displayed the fact that information aggregators meet consistent economic behaviors for fairly general settings, and that these behaviors may be extracted using simple derivations.

Our results might have applications, and implications, on various economic areas whose theoretical and empirical results depend on the use of functions embodied in the LDAs studied in this paper. This includes, for instance, theories such as imperfect competition, international trade and growth theory. LDAs might make it possible to transfer, in the economic frameworks, three main advantages that Bregman divergences in general have extensively brought to statistics, classification and even geometry \cite{bmdgCWj,nbnOB,nnOW}. First, the analytical expression of a LDA can be extremely complex, yet the abstraction of the general form (\ref{bregmeangen}), and its properties (Theorems \ref{th1}, \ref{thagh}), makes it possible for fairly complex behaviors to be derived in a simple manner. Second, it has been shown in the aforementioned fields that algorithms for solving different problems on a single divergence could be generalized to solving the same problems on any divergence. Last, but not least, LDAs encode such a large variety of functions that they could make it easier to find tight fits to economic data.


\subsection*{Acknowledgments}

The authors thank Pierre Cahuc for having read an earlier draft of this paper. R. Nock and N. Sanz gratefully acknowledge support from the State Secretary of Ultraperipheric Regions through grant 07MAR15 ``Coordination Failures and Distortions in Economic Models''. F. Nielsen and R. Nock are supported by National Research Agency (ANR) Blanc project ANR-07-BLAN-0328-01 ``Computational Information Geometry and Applications''.

\vskip 0.2in
\bibliography{../../BIB/bibgen}
\bibliographystyle{plain}

\section{Appendix}

\subsection{Proof of Theorem \ref{th0}}

Theorem 4 in \cite{bgwOT} states the result for a function that meets [1], [3] and $F(\bm{x}, \bm{x}) = 0, \forall \bm{x} \in {\mathbb{R}}^d$ (the $\Leftarrow$ in [2]). $F$ also meets identity of indiscernibles as otherwise indeed, (\ref{defmin}) would be violated for any distribution with distinct $\bm{x}$ and $\bm{y}$ as support and such that $F(\bm{x}, \bm{y}) = 0$. In this case, the solution of the left-hand side of (\ref{defmin}) would be $\bm{x}$ or $\bm{y}$, but not their average. 

\subsection{Proof of Theorem \ref{th1}}
 Because of its independent interest for Bregman divergences, we state the proof in the most general form: we do not make the assumption that the generator is separable (\ref{defsep}). Without loss of generality, we assume $\Gamma=1$ in (\ref{bregmeangen}). The concavity of $\mu_\varphi$ means:
\begin{eqnarray}
\es_j \bm{\nabla}^{-1}_\varphi(\es_i \bm{\nabla}_\varphi(\bm{x}_{ij})) & \leq & \bm{\nabla}_\varphi^{-1}(\es_i \bm{\nabla}_\varphi(\es_j \bm{x}_{ij})) \:\:. \label{defconc}
\end{eqnarray}
Let $\bm{x}_{ij} \defeq \bm{\nabla}^{-1}_\varphi (\bm{x}'_{ij})$ for $\bm{x}'_{ij} \in \mathrm{im}(\bm{\nabla}_\varphi)$. Applying $\bm{\nabla}_\varphi$ on both sides ($\varphi$ is strictly convex, so $\bm{\nabla}_\varphi$ is bijective) and replacing yields:
\begin{eqnarray}
\bm{\nabla}_\varphi(\es_j \bm{\nabla}^{-1}_\varphi(\es_i \bm{x}'_{ij})) & \leq & \es_i \bm{\nabla}_\varphi(\es_j \bm{\nabla}^{-1}_\varphi (\bm{x}'_{ij})) \:\:. \label{defconv}
\end{eqnarray}
Eq. (\ref{defconv}) states the convexity of the LDA $\bm{\mu} \defeq \bm{\nabla}_\varphi(\es_{{\mathcal{D}}} \bm{\nabla}^{-1}_\varphi (\bm{X}))$, but Legendre duality implies $\bm{\nabla}_\varphi = \bm{\nabla}^{-1}_{\varphi^\star}$, and we get $\mu = \mu_{\varphi^\star}$, the dual of LDA $\mu_\varphi$. The proof starting from the convexity of $\mu_\varphi$ follows the same path.

\subsection{Proof of Theorem \ref{thagh}}

We prove the first part of the Theorem relating $\mu_\varphi$ and $\bm{s}$. Without loss of generality and to save notations, we make the proof for separable generators (\ref{defsep}) and take a route slightly different from a direct use of Jensen's inequality. Furthermore, we make the proof for the concave case, assuming (\ref{bregmeangen}) holds for $\mu_\varphi$. Division by $\Gamma$ does not change the regime, and so $\mu_{\varphi} / \Gamma =  \nabla^{-1}_{\phi}\left( \sum_{i=1}^{m} {(\gamma_i/\Gamma) \nabla_{\phi}(x_i)} \right)$ is concave. We compute the tangential hyperplane to $\mu_{\varphi}/\Gamma$ when $x_i = \tilde{x} \in \mathrm{int}(\mathrm{dom}(\phi)), \forall i = 1, 2, ..., m$. We know that since $\mu_\varphi/\Gamma$ is concave, it shall be located below this hyperplane. We have $\bm{\nabla}_{\mu_{\varphi}/\Gamma} = [\cdots (\gamma_i/\Gamma) \phi^{[2]}(x_i) / \phi^{[2]}(\mu_{\varphi}) \cdots]^\top$, and so the tangential hyperplane to $\tilde{\bm{x}} = [\cdots \tilde{x} \cdots]^\top$ on any $\bm{x}$ is (with $\kappa_i \defeq \gamma_i / \Gamma$):
\begin{eqnarray}
z(\bm{x}) & = & \mu_{\varphi}(\tilde{\bm{x}}) + \langle \bm{x} - \tilde{\bm{x}}, \bm{\nabla}_{\mu_{\varphi}/\Gamma}(\tilde{\bm{x}}) \rangle \nonumber\\
 & = & \tilde{x} + \langle \bm{\kappa}, \bm{x} \rangle - \tilde{x} \nonumber\\
 & = & \langle \bm{\kappa}, \bm{x} \rangle = \sum_{i=1}^{m} {(\gamma_i/\Gamma) x_i}\:\:.
\end{eqnarray}
We obtain $\mu_{\varphi} / \Gamma \leq \sum_{i=1}^{m} {(\gamma_i/\Gamma) x_i}$. This yields to the statement that $\mu_{\varphi} \leq \bm{s}$, as claimed. The proof for the convex case is similar. The second part of the Theorem is an immediate consequence of the first part, and so this ends the proof of Theorem \ref{thagh}.

\subsection{Proof of Lemma \ref{lemcstar}}

We have $\phi^{[1]}_{\mathrm{\textsc{ces}}}(x) = a(2\sigma-1) x^{(\sigma-1)/\sigma} /\sigma$ and $\left(\phi^{[1]}_{\mathrm{\textsc{ces}}}\right)^{-1}(x) = (\sigma x / (a (2\sigma-1)))^{\sigma/(\sigma-1)}$. There remains to remark that $\Beta = \sum_{i=1}^{m} {\beta_i} = \Beta^{\frac{\sigma}{\sigma-1}}$, use (\ref{bregmeangen}) and get:
\begin{eqnarray}
\mu_{\phi_{\mathrm{\textsc{ces}}}} & = & \Beta^{\frac{\sigma}{\sigma-1}} \left( \frac{1}{\Beta^{\frac{\sigma}{\sigma-1}}} \sum_{i=1}^{m} {\Beta^{\frac{1}{\sigma-1}} \beta_i x_i^{\frac{\sigma-1}{\sigma}}} \right)^{\frac{\sigma}{\sigma-1}}\:\:,
\end{eqnarray}
which, after simplification, gives (\ref{defgences}), as claimed.

\subsection{Proof of Theorem \ref{fundth}}

Without loss of generality, we first show that, if $x_\star$ is a LDA that meets (\ref{fixg}), it is a CES. We thus assume the form (\ref{bregmeangen}) for $x_\star$. Using (\ref{bregmeangen}), we obtain that $x_\star$ satisfies (we replace in this proof notation $\phi_x$ by the simpler $\phi$ for the sake of readability):
\begin{eqnarray}
\phi^{[1]}\left(\frac{x_\star}{\Gamma}\right) & = & \frac{1}{\Gamma} \sum_{i=1}^{m} {\gamma_i \phi^{[1]}(x_i)} \label{propc1}\:\:,
\end{eqnarray}
with  $\gamma_i \geq 0,  \forall i =1, 2, ..., m$, and $\Gamma \defeq \sum_{i=1}^{m} {\gamma_i}$. If we differentiate (\ref{fixp}) with respect to any $x_i$, using (\ref{bregmeangen}), we get:
\begin{eqnarray}
\frac{z_i}{z_\star} & = & \frac{\gamma_i\phi^{[2]}(x_i)}{\phi^{[2]}(x_\star / \Gamma)}\:\:. \label{genpc}
\end{eqnarray}
We multiply both sides by $x_i / (\Gamma x_\star)$, sum for all $i$, simplify via (\ref{fixp}), rearrange, and get:
\begin{eqnarray}
\frac{x_\star}{\Gamma} \phi^{[2]}\left(\frac{x_\star}{\Gamma}\right) & = & \frac{1}{\Gamma} \sum_{i=1}^{m} {\gamma_i x_i \phi^{[2]}(x_i)} \:\:. \label{propc2}
\end{eqnarray}
Now, we match (\ref{propc1}) with (\ref{propc2}), and get that $\phi$ \textit{must} satisfy:
\begin{eqnarray}
\exists \kappa \in {\mathbb{R}}_* \mbox{ s.t. } \phi^{[1]}(x) & = & \kappa x \phi^{[2]}(x), \forall x \in \mathrm{dom}\phi\:\:. \label{superdom}
\end{eqnarray}
The solution is found to be $\phi^{[1]}(x) \propto x^\kappa$, i.e.:
\begin{eqnarray}
\phi(x) & = & \frac{d}{\kappa+1} x^{\kappa+1} \:\:, \label{genphic}
\end{eqnarray} 
with $d \in {\mathbb{R}}_*$ any constant that keeps (\ref{genphic}) convex. Matching (\ref{genphic}) with (\ref{propnces}) implies $\sigma = 1/(1-\kappa)$, and we get the proof that $x_\star$ is a CES.\\

Lemma \ref{lempstar} then implies that $z_\star$ is also a CES. The proof that $\phi^{[2]}_z = d \left(\phi^{[2]}\right)^{-1}$ for some $d \in {\mathbb{R}}_*$ follows from the expressions of $\phi_c$ and $\phi_p$ in Lemmata \ref{lemcstar} and \ref{lempstar}.

\subsection{Proof of Theorem \ref{constelast}}

It is well-known that the property is true for any CES, so we investigate the reverse implication, and given any LDA $x_\star$, let $\phi$ (which replaces $\phi_x$ for the sake of readability) denote its generator. Using (\ref{bregmeangen}), we obtain:
\begin{eqnarray}
{\mathrm{\textsc{e}}}_{x_\star}^{x_i} & = & \frac{\gamma_i x_i \phi^{[2]} (x_i)}{x_\star \phi^{[2]} (x_\star / \Gamma)} \:\:.
\end{eqnarray}
Summing this for $i=1, 2, ..., m$ and rearranging, we get that the sum of elasticities of $x_\star$ is one implies:
\begin{eqnarray}
\frac{x_\star}{\Gamma} \phi^{[2]} \left(\frac{x_\star}{\Gamma}\right) & = & \frac{1}{\Gamma}\sum_{i=1}^{m} {\gamma_i x_i \phi^{[2]} (x_i)}\:\:.
\end{eqnarray}
This is exactly (\ref{propc2}). There remains to conclude that $x_\star$ is a CES as in the proof of Theorem \ref{fundth}.

\subsection{Proof of Theorem \ref{constsubs}}

Implication $\Leftarrow$ is folklore, so we investigate the reverse implication. For any LDA $x_\star$ whose generator is denoted $\phi$, ${\mathrm{\textsc{e}}}_{x_\star}^{x_i \rightarrow x_j} = 1$ implies, with $\tilde{x}_{ij} \defeq x_j / x_i$:
\begin{eqnarray}
\frac{\mathrm{d}{\mathrm{\textsc{s}}}_{x_\star}^{x_i \rightarrow x_j}}{\mathrm{d} \tilde{x}_{ij}} & = & \frac{{\mathrm{\textsc{s}}}_{x_\star}^{x_i \rightarrow x_j}}{\tilde{x}_{ij}}\:\:,
\end{eqnarray}
implying $\mathrm{\textsc{s}}_{x_\star}^{x_i \rightarrow x_j} = \kappa \tilde{x}_{ij}$, for some $\kappa > 0$ which does not depend on $x_i$ or $x_j$. We obtain that $x_\star$ satisfies the following PDE:
\begin{eqnarray}
x_i \left( \frac{\partial x_\star}{\partial x_i} \right) - \kappa x_j \left( \frac{\partial x_\star}{\partial x_j} \right) & = & 0\:\:. \label{pyy}
\end{eqnarray}
Because $x_\star$ is a LDA, we have $\partial x_\star / \partial x_i = \gamma_i \phi^{[2]}(x_i) / \phi^{[2]}(x_\star)$, and so (\ref{pyy}) becomes:
\begin{eqnarray}
\gamma_i x_i \phi^{[2]}(x_i) & = & \kappa \gamma_j x_j \phi^{[2]}(x_j)\:\:. \label{defkkk}
\end{eqnarray}
Since $\kappa, \gamma_i, \gamma_j > 0$ and (\ref{defkkk}) holds for any $x_i, x_j \in \mathrm{int}(\mathrm{dom}\phi^{[1]})$, we obtain that $x \phi^{[2]}(x)$ is constant. This yields $\phi(x) = b (x\log x - x)$ for some constant $b \in {\mathbb{R}}_{+,*}$, the generator of Cobb-Douglas LDA (\ref{cobb}), as claimed.

\textit{Remark}: (\ref{defkkk}) also proves that if $m>2$ \textit{and} we require unit substitution elasticity between more than two goods, then necessarily $\kappa = 1$ and $\gamma_i = \gamma_j$, $\forall i, j= 1, 2, ..., m$. By means of words, fixing unit substitution elasticity for more than two goods implies equal weights for the goods. Clearly, this is a property of Cobb-Douglas aggregator rather than a restriction for LDAs.

\subsection{Proof of Theorem \ref{consthom}}

Implication $\Leftarrow$ is folklore, so we investigate the reverse implication. For any LDA $x_\star$ whose generator is denoted $\phi$, (\ref{defhom}) implies:
\begin{eqnarray}
\Gamma \nabla^{-1}_{\phi} \left(\frac{1}{\Gamma} \sum_{i=1}^{m} {\gamma_i \nabla_\phi(\lambda x_i)}\right) & = & \lambda^a \Gamma \nabla^{-1}_{\phi} \left(\frac{1}{\Gamma} \sum_{i=1}^{m} {\gamma_i \nabla_\phi(x_i)}\right)\:\:.
\end{eqnarray}
Take some $x_i$, $i=1, 2, ..., m$, and differentiate both sides in $x_i$. We get after simplification:
\begin{eqnarray}
\frac{\lambda \phi^{[2]}(\lambda x_i)}{\phi^{[2]}(\lambda^a x_\star)} & = & \frac{\lambda^a \phi^{[2]}(x_i)}{\phi^{[2]}(x_\star)} \:\:. \label{defc1}
\end{eqnarray}
\begin{itemize}
\item Case 1: $a\neq 1$. Suppose that $x_i = x_j = z / \lambda, \forall i,j = 1, 2, ..., m$, which implies $x_\star = z / \lambda$ as well. Eq. (\ref{defc1}) simplifies to:
\begin{eqnarray}
\phi^{[2]}(\lambda^{a-1} z) & = & \frac{1}{\lambda^{a-1}} \phi^{[2]}(z)  \:\:,
\end{eqnarray}
i.e. $\phi^{[2]}$ is homogeneous of degree $-1$. Euler's homogeneous function Theorem implies that $\phi^{[2]}$ satisfies the following PDE:
\begin{eqnarray}
x \phi^{[3]} (x) + \phi^{[2]}(x) & = & 0\:\:,
\end{eqnarray}
whose solution is $\phi^{[2]}(x) \propto 1/x$ (with a positive factor), i.e. $\phi(x) = b (x\log x - x)$ for some constant $b \in {\mathbb{R}}_{+,*}$, the generator of Cobb-Douglas LDA (\ref{cobb}), as claimed.
\item Case 2: $a=1$. In this case, (\ref{defc1}) implies:
\begin{eqnarray}
\phi^{[2]} (\lambda x) & = & g(\lambda) \phi^{[2]}(x)\:\:, \label{propg}
\end{eqnarray} 
for any function $g(\lambda) \in {\mathbb{R}}_*$. Suppose without loss of generality that $g$ is $C_1$, so that we can take the route of the proof of Euler's homogeneous function Theorem. We differentiate (\ref{propg}) in $\lambda$, and take the resulting equation for $\lambda = 1$. We obtain the following PDE:
\begin{eqnarray}
x \phi^{[3]} (x) - g^{[1]}(1) \phi^{[2]} (x) & = & 0 \:\:,
\end{eqnarray}
i.e. $\phi^{[2]}(x) \propto x^\kappa$, where $\kappa \in {\mathbb{R}}_*$ is some constant. We obtain that $\phi$ is either of the form of (\ref{propnces}), or (\ref{cobb}), the generators of CES and Cobb-Douglas LDAs, as claimed.
\end{itemize}

\subsection{Proof of Lemma \ref{mstlda}}

After differentiation on some $x_i$, should it be a LDA, any MST aggregator $x_\star$ with generator $\phi$ would satisfy:
\begin{eqnarray}
\gamma_i \times \frac{-\theta\exp(\theta \gamma_i x_i)}{(1 - \exp(\theta \gamma_i x_i))} \times x_\star & = & \gamma'_i \times \phi^{[2]} (x_i) \times \frac{1}{\phi^{[2]} (x_\star)}\:\:, \label{condmst}
\end{eqnarray}
with $\gamma'_i$ the LDA weight for $x_i$. This would imply $\phi^{[2]} (x) = 1 / x$, from which the simplification of (\ref{condmst}) yields that regardless of the value of $x_i$, the corresponding weights $\gamma_i$ and $\gamma'_i$ must satisfy $- \theta \gamma_i x_i \exp(\theta \gamma_i x_i) = \gamma'_i (1 - \exp(\theta \gamma_i x_i))$, impossible.

\subsection{Proof of Theorem \ref{thutil}}

Let $\lambda$ be the Lagrange multiplier for (\ref{fixu}), so that the Lagrangian is $ L \defeq u_\star + \lambda ( r - p_\star c_\star - m )$, and we obtain the following stationarity conditions for the optimum:
\begin{eqnarray}
\frac{\partial L}{\partial c_\star} & = & \frac{\gamma \phi^{[2]}_u(c_\star)}{\phi^{[2]}_u (u_\star)} - \lambda p_\star= 0\:\:, \label{firste}\\
\frac{\partial L}{\partial m} & = & \frac{(1-\gamma) \phi^{[2]}_u (m/p_\star)}{p_\star \phi^{[2]}_u(u_\star)} - \lambda= 0\:\:. \label{seconde}
\end{eqnarray}
Solving (\ref{seconde}) for $\lambda$ and simplifying (\ref{firste}) yields:
\begin{eqnarray}
\phi^{[2]}_u(c_\star) & = & \frac{1-\gamma}{\gamma} \phi^{[2]}_u \left(\frac{m}{p_\star} \right) \:\:.
\end{eqnarray}
There remains to use the identity $p_\star c_\star + m = r$ to get (\ref{eqcs}) and (\ref{eqm}), as claimed.

\end{document}